\documentclass[aps,prl,twocolumn,superscriptaddress,showpacs]{revtex4}
\usepackage{times,xspace}
\usepackage{amsbsy,amssymb,amsmath,bm}
\usepackage{amsfonts}
\usepackage{amsmath}
\usepackage{amssymb}
\usepackage[final]{graphicx}
\begin{document}
\title{Radiation Due to Josephson Oscillations in Layered Superconductors}
\author{L.N. Bulaevskii}
\affiliation{Los Alamos National Laboratory, Los Alamos, New Mexico 87545}
\author{A.E. Koshelev}
\affiliation{Materials Science Division, Argonne National Laboratory, Argonne, Illinois
60439 }
\date{\today}

\begin{abstract}
We derive the power of direct radiation into free space induced by
Josephson oscillations in intrinsic Josephson junctions of highly
anisotropic layered superconductors. We consider the super-radiation
regime for a crystal cut in the form of a thin slice parallel to the
$c$-axis. We find that the radiation correction to the
current-voltage characteristic in this regime depends only on
crystal shape. We show that at large enough number of junctions
oscillations are synchronized providing high radiation power and
efficiency in the THz frequency range. We discuss crystal parameters
and bias current optimal for radiation power and crystal cooling.
\end{abstract}

\maketitle

Josephson junctions (JJs), as sources of tunable continuous
electromagnetic radiation, were discussed for a long time after the
prediction of the ac Josephson effect.\cite{Jo} Early measurements
\cite{Lan} demonstrated that emittence from a single JJ has very low
power, typically $\sim10^{-6}$ $\mu$W. Since then a significant
effort has been devoted to develop JJ arrays as coherent sources of
radiation, see, e.g., Refs.~\onlinecite{Jain,Lukens}. A major
challenge is to force all JJs in array to emit coherently, so that
power increases proportionally to the square of the total number of
junctions.\cite{Jain} In particular, for an array of 500 junctions a
maximum power of the order of 10 $\mu$W at discrete frequencies
$\leq0.4$ THz has been achieved so far in the super-radiation
regime.\cite{Han} The difficulties to synchronize many artificial
JJs are related mainly to the facts that artificial junctions always
have slightly different parameters, especially the Josephson
critical current, and that one cannot put many of them at distances
smaller than a wavelength but needs to distribute them over a
wavelength or more.\cite{Lukens,Jain} Also, as the maximum frequency
is limited by the superconducting gap, it can not exceed several
hundred gigahertz for structures fabricated out of conventional
superconductors.

Layered high-temperature superconductors like
Bi$_{2}$Sr$_{2}$CaCu$_{2}$O$_{8}$ (BSCCO) offer a very attractive
alternative for developing radiation
sources.\cite{KoySSC95,Kleiner,Lat,Tach} A large value of the gap
(up to 60 meV) allows for very high Josephson frequencies, which can
be brought into the practically important terahertz range. Moreover,
intrinsic JJs (IJJS) have much closer parameters than artificial
ones as these parameters are controlled by the atomic crystal
structure rather than by amorphous dielectric layer in artificial
JJs. Also, layered superconductors provide a very high density of
IJJs (1 per 15.6 \AA \ along the $c$-axis) and thus it is easy to
reach super-radiation regime with many junctions on the scale of
radiation wave length. In this regime the radiated electromagnetic
field effectively couples JJs and helps to synchronize them. In this
Letter we demonstrate that the superradiation regime indeed results
in the synchronization of Josephson oscillations in IJJs in zero
external magnetic field. Thus $c$-axis current biased crystal may
work as source of Josephson coherent emission of radiation (JOCER).
We calculate the radiation power and IV characteristics and discuss
an optimal crystal geometry accounting for heating due to
quasiparticle dissipation.

So far mostly radiation from the flux flow of the Josephson vortices
was discussed in the literature.\cite{KoySSC95,Kleiner,Lat,Tach} The
inductive interlayer coupling typically promotes formation of the
\emph{triangular vortex lattice}. However, to generate noticeable
outside radiation oscillations induced by the moving lattice have to
be in phase in different layers, which is realized only if the
moving vortices form a \emph{rectangular lattice}. No regular way to
prepare such a lattice is known at present. In addition, it seems to
be unstable in most of parameter space.\cite{Art,ak} Here we
consider the synchronization of the Josephson oscillations by
radiation field in the simplest case, when dc magnetic field is not
applied and only radiation itself introduces the in-plane phase
gradients.

\begin{figure}[ptb]
\begin{center}
\includegraphics[width=0.45\textwidth]{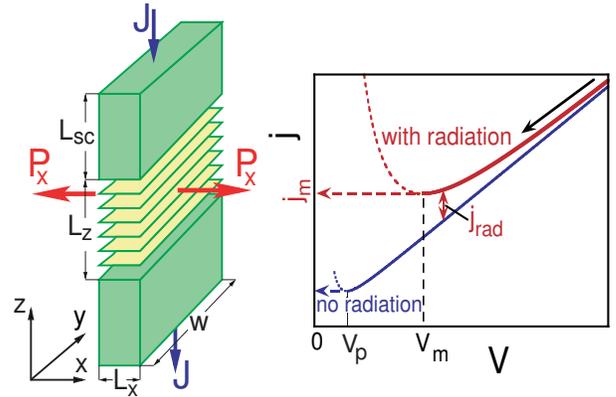}
\end{center}
\caption{(color online) \emph{Left:} Schematic picture of layered
superconductor placed in between plate-like leads serving as
screens. The directions of the dc transport current $J$, and of the
radiation Poynting vectors $P_{x}$ are shown. \emph{Right:}
Influence of radiation on current-descending branch of the IV
dependence.} \label{SchemFig}
\end{figure}

In the resistive state phases $\varphi_{n}$ oscillate at the
Josephson frequency, $\omega_{J}=2eV/\hbar$, where the voltage $V$
between the neighboring layers is induced by interlayer dc current.
For uniform oscillations and identical junctions the voltage $V$ is
the same in all junctions (except, possibly, top and bottom
junctions in the stack) because the same current flows between all
layers. To reach a resistive state the pulse of dc current exceeding
the Josephson critical current should be applied but then current
may be diminished to reach necessary voltage. Transport measurements
in BSCCO have shown that the resistive state on current-descending
branch is preserved down to the voltage $V_p\sim \hbar\omega_p/2e$,
where  $\omega_{p}$, is the Josephson plasma frequency see, e.g.,
Ref.\ \onlinecite{LYB},
 [$\omega_{p}/(2\pi)\approx0.15$ THz in optimally doped BSCCO].
We consider crystal with sizes $L_{x}$, $w$ and $L_{z}=Ns$ in the
directions $x$,$y$, and $z$ respectively, see Fig.~\ref{SchemFig}.
The conditions $L_{x} ,L_{z}\lesssim{k}_{\omega}^{-1}=c/\omega_{J}$,
are necessary for the super-radiation regime. The boundary
conditions for the oscillating phase are sensitive to distribution
of outside em fields which, in turn, depend on geometry of the stack
and electric contacts. We consider the simplest geometry assuming
that (i) $w\gg{k}_{\omega}^{-1}$ so that all quantities are $y$
independent, and (ii) JJ stack is bounded by metallic or
superconducting contacts with the same lateral sizes as the stack
which extend in the $z$ direction over distance
$L_{sc}\gg{k}_{\omega}^{-1}$ (see Fig.\ \ref{SchemFig}). We also
assume that the contact material has a very small surface impedance
so that the ac electric field at the contact surface is negligible.
Such contacts serve as ``screens'', restricting radiation to
half-infinite spaces $|x|> L_{x}/2$. This greatly simplifies the
analytical consideration. As dissipation increases with $L_{x}$ and
radiation is not, we argue that optimal crystal should be platelike
with $L_{x}< L_{z}$.

To find the phase differences $\varphi_{n}(x,t)$ inside the crystal we solve
finite-difference differential equations, see, e.g., Ref.~\onlinecite{ak},
\begin{align}
& \frac{\partial^{2}\varphi_{n}}{\partial\tau^{2}}=(\alpha\nabla_{n}
^{2}\!-1)\left(  \nu_{c} \frac{\partial\varphi_{n}}{\partial\tau}\!+
\sin \varphi_{n}\!-\nabla_u h_{y,n}\right) \label{Phase_eq},
\\
&  \left(  \nabla_{n}^{2}-\ell^{-2}\hat{T}_{ab}\right) h_{y,n}+\hat
{T}_{ab}~\nabla_u\varphi_{n}=0. \label{h_eq}
\end{align}
We use reduced $x$ coordinate, $u\!=\!x/\lambda_{J} $ normalized to
$\lambda_{J}\!=\!\gamma s$, reduced time, $\tau\!=\!\omega_{p}t$ and
$\omega\!=\!\omega_{J}/\omega_{p}$, with
$\omega_{p}\!=\!c/(\lambda_{c}\sqrt{\epsilon_{c}})$, and reduced
magnetic field $h_{y,n}=B_{y,n}/B_{c}$ with
$B_{c}\!=\!\Phi_{0}/(2\pi \lambda_{ab}\lambda_{c})$, where $B_{y,n}$
is the magnetic field between the layers $n$ and $n\!+\!1$. Here
$\epsilon_{c}$ is the $c$-axis
%SHORTENING
%high-frequency
dielectric constant
inside the superconductor, $\lambda_{ab}$ and $\lambda_{c}$ are the
London penetration lengths, $\gamma=\lambda_{c}/\lambda_{ab}$ is the
anisotropy ratio. In terms of these parameters the Josephson
critical current is $J_{c} =\Phi_{0}c/(8\pi^{2}s\lambda_{c}^{2})$.
Further, $\hat{T}_{ab}\!\equiv\!
1\!+\!\nu_{ab}\partial/\partial\tau$, $\ell\equiv\lambda_{ab}/s$ ,
$\nabla_{n}^{2}$ notates the discrete second derivative operator,
$\nabla_{n} ^{2}A_{n}\!=\!A_{n\!+\!1}+A_{n\!-\!1}-2A_{n} $, and
$\alpha\!\sim\!0.1\!-\!1$ is the parameter of the capacitive
coupling.\cite{Koyama} The dissipation parameters, $\nu _{ab}\!
=\!4\pi\sigma_{ab}/(\gamma^{2}\epsilon_{c}\omega_{p})$ and $\nu
_{c}\!=\!4\pi\sigma_{c}/(\epsilon_{c}\omega_{p})$, are determined by
the quasiparticle conductivities, $\sigma_{ab}$ and $\sigma_{c}$,
along and perpendicular to the layers, respectively.
%SHORTENING
%Eqs.~(\ref{Phase_eq}) and (\ref{h_eq}) represent just charge
%conservation law accounting for inductive and capacitive coupling of
%intrinsic JJs.
The electric field inside the superconductor between
the layers $n$ and $n+1$
%SHORTENING
%in terms of the phase difference
is given by
\begin{equation}
(1-\alpha\nabla_{n}^{2})E_{zn}=(B_{c}\ell/\sqrt{\epsilon_{c}}
)(\partial\varphi_{n}/\partial\tau). \label{e}
\end{equation}
The typical parameters of optimally doped BSCCO at low temperatures
are $\epsilon_{c}\!=\!12$, $s\!=\!15.6$ \AA , $\gamma\!=\!500$,
$\lambda_{ab}\!=\!200$ nm, $J_{c}\!=\!1700$ A/cm$^{2}$,
$\sigma_{c}(0)\!=\!2\cdot10^{-3}$ (ohm$\cdot$cm)$^{-1}$,
$\sigma_{ab}(0)\!=\!4\cdot10^{4}$ (ohm$\cdot$cm)$^{-1}$.\cite{LatKB}
This gives $\ell\approx 130$, $\nu_{ab} \approx0.2$, and
$\nu_{c}\approx 2\cdot10^{-3}$. An important feature of BSCCO is
higher in-plane dissipation in comparison with $c$ axis one.

The boundary conditions, i.e., relations between time and space
derivatives of $\varphi_{n}$ at the edges parallel to $(y,z)$, are
determined by the relations between the electric and magnetic fields
in the outside media. As the $y$ and $z$ sizes of the system
(crystal and screens) are assumed to be larger than the wavelength,
the dielectric media can be treated as infinite in these directions.
Such a half-infinite space geometry allows us to find the boundary
conditions analytically. From the Maxwell equations in the free
space we find relation between the magnetic,
$\mathbf{B}\!=\!(0,B_{y},0)$, and the electric,
$\mathbf{E}\!=\!(E_{x},0,E_{z})$, fields at the boundaries. We
assume that there are only outgoing waves from the crystal $(y,z)$
edges meaning that the fields have the coordinate and time
dependence $\exp(ik_{x}|x|\!+\!ik_{z}z\!-\!i\omega\tau)$, where
$k_{x}\!=\!\mathrm{sign}(\omega)({k}_{\omega}
^{2}\!-\!k_{z}^{2})^{1/2}$ for $k_{z}^2<{k}_{\omega}^2$ and
$k_{x}=i(k_{z} ^{2}\!-\!{k}_{\omega}^{2})^{1/2}$ for
$k_{z}^{2}\!>\!{k}_{\omega}^{2}$. The relations between fields at
$u\!=\!\pm\tilde{L}_{x}/2$~~ ($\tilde{L}_{x}
\!=\!L_{x}/\lambda_{J}$) are \cite{VL}
\begin{align}
&  B_{y}(\omega,{k}_{z})=\mp\zeta_{\omega}(k_{z})E_{z}(\omega,k_{z}
),\label{BoutEout}\\
&  \zeta_{\omega}(k_{z})=
\genfrac{\{}{.}{0pt}{}{|{k}_{\omega}|({k}_{\omega}^{2}-k_{z}^{2}
)^{-1/2}\text{, for
}k_{z}^2<{k}_{\omega}^2,}{-i{k}_{\omega}(k_{z}^{2}
-{k}_{\omega}^{2})^{-1/2}\text{, for }k_{z}^2>k_{\omega
}^2. }\nonumber
\end{align}
Inverse Fourier transform with respect to $k_{z}$ gives nonlocal
relation between the magnetic and electric fields at the edges. As
we assume that the screen material has small surface impedance, we
can neglect the electric field at $|z|>L_{z}/2$ and, using Eq.\
(\ref{e}), we obtain the reduced boundary condition connecting the
magnetic field $h_{y,n}$ with the phases at the edges ($h_{y,n}$ is
expressed via $\nabla\varphi_{n}$ by Eq.~(\ref{h_eq})),
\begin{align}
\pm h_{y,n,\omega}  &  =\frac{is\ell\omega}{2\sqrt{\epsilon_{c}}}
\sum_{m}(1-\alpha\nabla
_{m}^{2})^{-1}\varphi_{m,\omega}\nonumber\\
&  \times\left[  |{k}_{\omega}|J_{0}({k}_{\omega}s|n\!-\!m|)\!+\!i{k}_{\omega
}N_{0}({k}_{\omega}s|n\!-\!m|)\right]  , \label{BoundCond}
\end{align}
where $J_{0}(x)$ and $N_{0}(x)$ are the Bessel functions.

We consider high-frequency Josephson oscillations,
$\omega=\omega_{J} /\omega_{p}\gg1$, in the case of layered crystals
with large number of junctions $N\gtrsim\ell$. This allows us to
neglect finite-size effects along the $z$-axis. The equation for
uniform solution $\varphi_{n}(u,\tau )=\varphi(u,\tau)$ is
\begin{equation}
\partial^{2}\varphi/\partial\tau^{2}+\nu_{c}\partial\varphi/\partial\tau
+\sin\varphi-\ell^{2}\nabla_{u}^{2}\varphi=0.
\end{equation}
In the limit $\omega\gg1$ we look for the solution in the form
$\varphi (u,\tau)=\omega\tau+\phi(u,\tau)$ with $\phi\ll1.$
Eq.~(\ref{BoundCond}) gives the boundary conditions for $\phi$ at
$u=\pm\tilde{L}_{x}/2$ \cite{VL}
\begin{align}
&  \nabla_{u}\phi=\pm i\omega\zeta~\phi,\\
&
\zeta=\frac{L_{z}}{2\ell\sqrt{\epsilon_{c}}}[|k_{\omega}|-ik_{\omega
}\mathcal{L}_{\omega}],\ \ \ \ \ \
\mathcal{L}_{\omega}\approx\frac{2}{\pi}\ln\left[
\frac{5.03}{|k_{\omega}|L_{z}}\right]. \nonumber
\end{align}
The solution is $\phi(u,\tau)=\operatorname{Im}\left[
\phi_{\omega}(u)\exp(-i\omega \tau)\right]  $, where
\begin{align}
\phi_{\omega}\!  &  =\!-(\omega^{2}+i\omega\nu_{c})^{-1}+A\cos\left(  \bar
{k}_{\omega}u\right) \hbox{ with } \bar{k}_{\omega}\!=\!\omega/\ell, \label{JO}\\
A\!  &  =\!i\zeta\{[\bar{k}_{\omega}\sin(\bar{k}_{\omega}\tilde{L}
_{x}/2)\!+\!i\zeta\omega\cos(\bar{k}_{\omega}\tilde{L}_{x}/2)](\omega
\!+\!i\nu_{c})\}^{-1}.\nonumber
\end{align}
Here $|\zeta| \ll1$. The first term in $\phi_{\omega}$ is the
amplitude of Josephson oscillations, while the second term describes
the electromagnetic waves propagating inside the junctions. They are
generated at the boundaries due to the radiation field.

Next we show that coherent radiation field, similar to all
junctions, in combination with intralayer dissipation stabilizes the
uniform Josephson oscillations. For that we have to consider a small
perturbation to the uniform solution, $\varphi_{n}
(u,\tau)\!=\!\omega\tau\!+\!\phi(u,\tau)\!+\!\vartheta_{n}(u,\tau)$
and verify that there is no perturbations, $\vartheta_{n}(u,\tau)$,
increasing with time. The general analysis is rather cumbersome.
Results in a closed form may be obtained only by use of
approximations valid in the limiting case considered here,
$L_x,L_z\ll k_{\omega}^{-1}$, $\omega\gg 1$, and $\ell\gg 1$.

Equations for $\vartheta_{n}(u,\tau)$ are
obtained by linearization of Eqs.\ (\ref{Phase_eq}) and
(\ref{h_eq}). The term
$\cos[\phi(\tau)]\vartheta_{n}(u,\tau)\!\approx
\!\cos(\omega\tau)\vartheta_{n}(u,\tau)$ in the linearized equation
couples harmonics with small frequency $\Omega$ with the
high-frequency terms $\Omega\pm\omega$. At $\omega\gg1$ we can
neglect coupling to the higher frequency harmonics $\Omega\pm
m\omega$ with $m>1$ and represent the phase perturbation (and field) as
\[
\vartheta_{n}\!\approx\! \sum_{q}\!\left[
\bar{\vartheta}_{q}\!+\!\sum_{\beta=\pm1}\!\tilde{\vartheta
}_{q,\beta}\exp(i\beta\omega\tau)\right]\sin(qn)\exp\left(
-i\Omega\tau\right)
\]
with $q=\pi k/(N\!+\!1)$, $k\!=\!1,2\dots,N$.
Here the complex
eigenfrequency $\Omega\!=\!\Omega(q)$ is assumed to be small,
$|\Omega|\ll\omega$, and has to be found. Stability means that
$\operatorname{Im}[\Omega]<0$ for all $q$. Substituting this
presentation into the linearized equations (\ref{Phase_eq}) and
(\ref{h_eq}), excluding oscillating magnetic fields, and separating
the fast and slow parts, we obtain coupled equations
\begin{align}
&  \left[  \frac{\Omega^{2}}{1\!+\!\alpha_{q}}+i\nu_{c}\Omega-\bar{C}\right]
\bar{\vartheta}_{q}+G_{q}^{-2}\nabla_{u}^{2}\bar{\vartheta}_{q}\!=\!\frac
{\tilde{\vartheta}_{q,+}\!+\!\tilde{\vartheta}_{q,-}}{2},\label{lc}\\
&  \left[  \frac{(\Omega\!\mp\!\omega)^{2}}{1\!+\!\alpha_{q}}\!+i\nu
_{c}(\Omega\!\mp\!\omega)\right]  \tilde{\vartheta}_{q,\pm}\!+\!G_{q,\pm}
^{-2}\nabla_{u}^{2}\tilde{\vartheta}_{q,\pm}\!=\!\frac{\bar{\vartheta}_{q}}
{2}.\label{fc}
\end{align}
Here
$\bar{C}\!\equiv\!\langle\cos\phi\rangle_{\tau}\!\approx\!\mathrm{Re}[\phi_{\omega}]/2$,
$\alpha_{q}\!\equiv\!\alpha\tilde{q}^{2}$ with $\tilde
{q}^{2}\!=\!2(1\!-\!\cos q)$,
$G_{q,\beta}^{2}\!=\!\tilde{q}^{2}/[1\!-\!i(\Omega-\beta\omega)\nu
_{ab}]\!+\!\ell^{-2}$, and $G_{q}=G_{q,0}$. Using Eqs.~(\ref{h_eq}),
(\ref{BoutEout}) (in the limit $k_{z}^2>k_{\omega}^2$ due to
$k_{\omega}L_z\ll \pi $), and (\ref{BoundCond}) we get the boundary
conditions for slow and fast components at $u=\pm\tilde{L}_{x}/2$
for $q\gg\pi/N$,
\begin{align}
\nabla_{u}\bar{\vartheta}_{q} &
=\!\pm\kappa_{0}\bar{\vartheta}_{q},\ \ \ \ \kappa _{0}\approx
G_{q}^{2}\Omega^{2}/[(1+\alpha_{q})\epsilon_{c}q\gamma
],\label{BC_slow}\\
\nabla_{u}\tilde{\vartheta}_{q,\beta} &
=\!\pm\kappa_{\beta}\tilde{\vartheta }_{q,\beta},\
\kappa_{\beta}\!\approx\!\frac{\left(  \Omega\!-\!\beta
\omega\right)
^{2}G_{q,\beta}^{2}}{(1\!+\!\alpha_{q})\epsilon_{c}q\gamma
}.\label{BC_fast}
\end{align}
Because of the condition $|\Omega|\ll\omega$, in most cases one can
neglect $\Omega$ in equation and boundary conditions for
$\tilde{\vartheta}_{q,\beta} $. We also assume
$\nu_{c}\ll1\ll\omega$ and neglect dissipation when it is not
essential. As $\bar{\vartheta}_{q}$ varies at the typical length
scale $\sim1/G_{q}\Omega$, which is much larger than $L_{x}$, the
coordinate-dependent part of $\bar{\vartheta}_{q}$ can be treated as
a small perturbation. Neglecting the coordinate dependence of
$\bar{\vartheta}_{q}$ in the equation for $\tilde{\vartheta}$, we
obtain the approximate solution of Eqs.~(\ref{fc}) and
(\ref{BC_fast}). Substituting it into Eq.\ (\ref{lc}), we obtain
Mathieu equation for the slow-varying component
\begin{equation}
\left(  \frac{\Omega^{2}}{1\!+\!\alpha_{q}}\!+\!i\nu_{c}\Omega\!-\!\frac
{\alpha_{q}}{2\omega^{2}}\!-\!V(u)\!+\!G_{q}^{-2}\nabla_{u}^{2}\right)
\bar{\vartheta}_{q}\!=\!0,\label{maineq}
\end{equation}
where the \textquotedblleft potential\textquotedblright\ is given by
$V(u)\!=\!V_{1}(u)\!+\!V_{2}(q,u)$,
\begin{align*}
V_{1}(u) &  \!\approx\!\frac{1}{2\omega^{2}}\operatorname{Re}\left[
\frac{i\zeta\omega\cos(\bar{k}_{\omega}u)}{\bar{k}_{\omega}\sin(\bar
{k}_{\omega}\tilde{L}_{x}/2)\!+\!i\zeta\omega\cos(\bar{k}_{\omega}\tilde
{L}_{x}/2)}\right]  ,\\
V_{2}(q,u) &
\!\approx\!\frac{1}{2\omega^{2}}\operatorname{Re}\left[
\frac{\kappa_{+}\cos
p_{+}u}{p_{+}\sin(p_{+}\tilde{L}_{x}/2)+\kappa_{+}
\cos(p_{+}\tilde{L}_{x}/2)}\right],
\end{align*}
and $p_{+}=\omega G_{q,+}$.
In the super-radiation regime,
$\bar{k}_{\omega}\tilde{L}_{x}\!=\!\omega \tilde{L}_{x}/\ell\ll1$,
in the lowest order with respect to $\omega\tilde {L}_{x}/\ell$,
the part $V_{1}(u)$ reduces to a constant, $V_{1}(u)\approx
{\cal K}_{\omega}/(2\omega^{2})$ with
${\cal K}_{\omega}=[{\cal L}_{\omega}({\cal L}_{\omega}+\varepsilon_ca)+1)
/[({\cal L}_{\omega}+\varepsilon_ca)^2+1]$
and $a\!=\!L_{x} /L_{z}$.

Equation (\ref{maineq}) and the boundary conditions (\ref{BC_slow})
determine the spectrum of small perturbations to the uniform
solution. Treating the coordinate-dependent part of
$\bar{\vartheta}_{q}$ as a small perturbation allows us to derive
the expression for $\Omega(q)$,
\begin{align}
\Omega^{2}\!+i\nu_{c}\Omega & \approx\left[
\alpha_{q}\!+{\cal K}_{\omega}-W_{2}(q)\right]  /\left(
2\omega^{2}\right),
\label{FreqResult}\\
W_{2}(q) & =\!\operatorname{Re}\left[2/[p_{+}\tilde{L}_{x}
(p_{+}/\kappa_{+}\!+\!\cot( p_{+}\tilde{L}_{x}/2)) ]\right].
\nonumber
\end{align}
From this result we can conclude that the main contribution to
stabilization of uniform oscillations comes from the term ${\cal
K}_{\omega}$, describing effective coupling of junctions due to the
radiation. Its stabilization effect increases with $L_z$ as ${\cal
K}_{\omega}\approx {\cal L}_{\omega}L_z/\epsilon_cL_x$ for
$L_z<\epsilon_cL_x$ and ${\cal K}_{\omega}\rightarrow 1$ for
$L_z>\epsilon_cL_x$. The charging-effect term, $\alpha_q$, also
contributes to stabilization. The term $W_{2}$ describes the effect
of modes $\tilde{\vartheta}_{q,\pm}$ induced inside the crystal due
to radiation. Formally, the $W_2$ term leads to instabilities
\emph{in the limit of zero dissipation} because its denominator
vanishes near the resonance values of $q$ given by $2(1\!-\!\cos
q)\!=\![ 2\pi m/(\omega\tilde{L}_{x})] ^{2}\gg1/\ell^{2}$, where $m$
is an integer. These instabilities correspond to parametric
excitation of the Fiske resonances described by Eq.~(\ref{fc}).
However, they are suppressed already by very small dissipation.
Indeed, at small dissipation we estimate the maximum value of
$W_{2}$ as
$[q\gamma\varepsilon_{c}\operatorname{Im}[p_+]\tilde{L}_{x}^{2}]^{-1}$,
where $\operatorname{Im}[p_{+}]\!
\approx\sin(q/2)\left(\nu_{c}\!+\!\nu_{ab}\omega^{2}\right)\!$. As
$\nu _{ab}\omega^{2}\gg\nu_{c}$, we see that $|W_{2,\max}|\!\ll \!
1$ for $\nu _{ab}\!\gg\!
1/(2\pi^{2}\gamma\varepsilon_{c})\!\sim\!10^{-5}$. For realistic
level of dissipation in BSCCO, $\nu_{ab}\!\sim 0.2$, the resonance
features in $W_{2}$ are completely washed out and $\left\vert
W_{2}\right\vert \ll1$ for all $q$'s. Thus the intralayer
dissipation stabilizes uniform oscillations but it does not affect
them in any other way.

We proceed now with derivation the radiation power and IV
characteristics in the super-radiation regime of uniform Josephson
oscillation. The Poynting vector $P_{x}$ at $x=\pm L_x/2$ in terms
of the oscillating phase is given by \cite{VL}
\[
P_{x}(\omega)\!=\!\pm\frac{\Phi_{0}^{2}\omega_{J}^{3}}{64\pi^{3}c^{2}sN}\sum
_{n,m}J_{0}({k}_{\omega}s|n\!-\!m|)| \varphi_{\omega} (\pm
L_x/2)|^{2},
\]
where the oscillating phase difference, $\varphi_{\omega}$, is
determined by Eq.~(\ref{JO}). Substituting this solution in the
limit $\bar{k}_{\omega}\tilde{L}_{x}\ll1$, we obtain  for the total
radiation power $\mathcal{P} _{\mathrm{rad}}(\omega)= P_{x}(\omega)L_{z}w$
going from one side,
\begin{align}
&  \mathcal{P}_{\mathrm{rad}}(\omega_J)/w\approx\lbrack\Phi_{0}^{2}\omega_{p}^{4}
N^{2}/(64\pi^{3}c^{2}\omega_{J})]\mathcal{L}(a),\nonumber\\
& \mathcal{L}(a)=a^{2}\epsilon_{c}^{2}/[(a\epsilon_{c}+\mathcal{L}_{\omega
})^{2}+1],
\end{align}
For small  $L_{x}$, $L_{x}\!\ll\! L_{z}/\epsilon_{c}$, $
\mathcal{P}_{\mathrm{rad}}\propto L_{x}^{2}$ and it is $N$-independent,
while for larger $L_{x}$ the geometrical factor
$\mathcal{L}(a) \rightarrow 1$, meaning that
$\mathcal{P}_{\mathrm{rad}}\propto N^{2}$ and it is $L_x$-independent.

The dc interlayer current density $J$ consists of quasiparticle
contribution, $\sigma_{c}V/s$, and the Josephson part
$J_{c}\sin\varphi$, averaged over time and coordinate. We derive for
$j=J/J_{c}$,
\begin{equation}
j=\nu_{c}\omega+ \frac{1}{2}\langle\mathrm{Im}[\phi_{\omega}] \rangle_{u} =
\nu_{c}\omega+\frac{\nu_{c}}{2 \omega^{3}}+\frac{\mathcal{L}(a)}{2\omega
^{2}\epsilon_{c}a}.
\end{equation}
The last term, $j_{{\rm rad}}$, describes contribution to the dc
current due to radiation losses. This part of the current multiplied
by the total voltage gives total radiation power
$2\mathcal{P}_{\mathrm{rad}}$. As a function of $\omega $ the
current $j$ has minimum at $\omega\!=\!
\omega_{\mathrm{m}}\!\approx\!
[\mathcal{L}(a)/\epsilon_{c}\nu_{c}a)]^{1/3}\!\gg \!1$. The
influence of radiation on the IV dependence is illustrated in Fig.\
\ref{SchemFig}. Only part of the IV characteristics at
$\omega\!>\!\omega_{\mathrm{m}}$ is stable. At the voltage
$V_{\mathrm{m}}\!=\!\hbar\omega_{\mathrm{m}}/2e$, corresponding to
the current $j_{\mathrm{m}}\!=\!(3/2)\nu_{c}\omega_{\mathrm{m}}$,
the stack jumps back to the static state. At this ``retrapping''
current the dissipation power is twice of the radiation power, i.e.,
for the conversion efficiency we obtain
$2\mathcal{P}_{\mathrm{rad}}/(\mathcal{P}_{\mathrm{dis}
}+2\mathcal{P}_{\mathrm{rad}})\leq1/3$.

An important issue is the stability of the coherent state with
respect to parameter variations from layer to layer, which may
include the crystal width, $\delta L_{xn}$, and the Josephson
current density, $\delta J_{cn}$. The coherent solution with the
same voltage drop in all junctions, $V$, must satisfy the current
conservation condition. Such solution can be built as
$\varphi_{n}(u,\tau)\!=\!\omega\tau\!+\!\phi_{\omega,n}(u,\tau)\!+\!\beta_{n}$,
where the additional phase shifts, $\beta_{n}$, compensate for
parameter variations. In the case of smooth parameter variations and
small charging parameter, we can derive correction to the local
current and obtain
\begin{equation}
\sin(\beta_{n})-\mathcal{S}\approx\frac{\delta L_{xn}}{\langle
L_{xn}\rangle}\frac{\nu_{c}\omega}{j_{\mathrm{rad}}}-\frac{\delta J_{cn}%
}{\langle J_{cn}\rangle}\mathcal{S},\nonumber
\end{equation}
with $\mathcal{S}=\left\langle \sin(\beta_{n})\right\rangle $ and
$\langle \dots \rangle$ means average over $n$. We can see that the
coherent state survives until $\delta L_{xn}/\langle
L_{xn}\rangle<j_{\mathrm{rad}}/(\nu _{c}\omega)$ and $\delta
J_{cn}<\langle J_{cn}\rangle$. These conditions do not impose too
demanding restrictions on the acceptable range of parameter
variations.

Evaluating the Joule heating, we find that the cooling rate $Q$ per
unit area of each crystal side $\parallel yz$ should be
\begin{equation}
Q\approx
\frac{\Phi_{0}^{2}\omega_{p}\nu_{c}}{32\pi^{3}\lambda_{c}^{2}s^{2}}
\omega^{2}L_{x}\approx  \left( \frac{\omega}{\omega_{\mathrm{m}}
}\right) ^{3} \frac{2\mathcal{P}_{\mathrm{rad}}}{L_{z}w}.
\label{heat}
\end{equation}
The maximum efficiency is reached at
$\omega\!\approx\!\omega_{\mathrm{m}}$. To achieve this at
frequencies close to 1 THz one needs to maximize $\omega
_{\mathrm{m}}\!\propto\![\mathcal{L}(a)/a]^{1/3}$ by optimizing the
crystal shape. At ${\cal L}_{\omega} \!\sim\!1$ we get
$\mathrm{max}[\mathcal{L}(a)/a]\!\approx\! \epsilon_{c}/4$ at
$a\!\approx\!1/\epsilon_{c}$. This gives $\omega_{\mathrm{m}
}\!\approx\! 5$ corresponding to $\omega_{J}/(2\pi)\!=\! 0.75$THz.
At this frequency, assuming $L_{z}\!=\!40$ $\mu$m, the optimum
lateral sizes are $L_{x}\!\approx\! 4$ $\mu$m and $w\!>\!300$
$\mu$m. Biased with the current density $\!\approx\!0.01J_{c} $,
such a crystal radiates with the power
$\mathcal{P}_{\mathrm{rad}}/w\!\approx\! 30$mW/cm from each side,
while it should be cooled with the rate $Q\!\approx\! 15$ W/cm$^{2}$
at each side.\cite{footnote} As $L_{x}$ increases,
$\mathcal{P}_{\mathrm{rad}}/w$ saturates to $\!\approx\!0.2$ W/cm at
$\omega \!\sim\! \omega_{m}\!\propto\! L_x^{-1/3}$, while $Q$
increases linearly with $L_{x}$. We note that increasing the number
of layers also promotes synchronization of oscillations by the
radiation field.

In conclusion, we have shown that uniform Josephson oscillations in
intrinsic junctions of layered superconductors are stable in the
superradiation regime at high frequencies $\omega_{J}\gg\omega_{p}$.
They lead to coherent radiation into free space with significant
power and efficiency as high as 1/3 at frequencies $\sim 1$ THz.
Important point is that to have reasonable cooling rate and strong
radiation the crystal should be in the form of thin plate along the
$c$ with large number of layers $N\sim 10^{4}$.

The authors thank M. Maley, I. Martin, K. Kadowaki, M. Tachiki, R.
Kleiner, A. Ustinov, and V. Kurin for numerous useful discussions.
This research was supported by the US DOE under the contracts \#
W-7405-Eng-36 (LANL) and \# DE-AC02-06CH11357 (ANL).


\begin{thebibliography}{99}
\bibitem{footnote}For effective cooling and mechanical stability the crystal may be
placed on the dielectric at one of $yz$ sides. The modification of
the boundary condition for this case is strightforward.
\end{thebibliography}
\end{document}